\documentclass{emulateapj}

\begin{document}
\slugcomment{Accepted to ApJ}

\title{ Hypervelocity Stars II.  The Bound Population}

\author{Warren R.\ Brown,
	Margaret J.\ Geller,
	Scott J.\ Kenyon,
	Michael J.\ Kurtz}

\affil{Smithsonian Astrophysical Observatory, 60 Garden St, Cambridge, MA 02138}
\email{wbrown@cfa.harvard.edu}

\and	\author{Benjamin C.\ Bromley}
\affil{Department of Physics, University of Utah, 115 S 1400 E, Rm 201, Salt Lake City, UT 84112}

\shorttitle{ Hypervelocity Stars II.  The Bound Population}
\shortauthors{Brown et al.}

\begin{abstract}

	Hypervelocity stars (HVSs) are stars ejected completely out of the Milky Way
by three-body interactions with the massive black hole in the Galactic center.  We
describe 643 new spectroscopic observations from our targeted survey for HVSs.  We
find a significant (3.5$\sigma$) excess of B-type stars with large velocities
$+275<v_{rf}<450$ km s$^{-1}$ and distances $d>10$ kpc that are most plausibly
explained as a new class of HVSs:  stars ejected from the Galactic center on bound
orbits.  If a Galactic center ejection origin is correct, the distribution of HVSs
on the sky should be anisotropic for a survey complete to a fixed limiting apparent
magnitude.  The unbound HVSs in our survey have a marginally anisotropic
distribution on the sky, consistent with the Galactic center ejection picture.

\end{abstract}

\keywords{
        Galaxy: halo ---
        Galaxy: center ---
        Galaxy: stellar content ---
        Galaxy: kinematics and dynamics ---
        stars: early-type
}

\section{INTRODUCTION}

	HVSs are a natural consequence of a massive black hole (MBH) in a dense
stellar environment like that in the Galactic center.  \citet{hills88} first pointed
out that a stellar binary encountering the Milky Way's central MBH can eject one
member of the binary as a HVS traveling at $\sim$1,000 km s$^{-1}$.  HVSs differ
from classical ``runaway'' stars because 1) HVSs are unbound and 2) the classical
supernova ejection \citep{blaauw61} and dynamical ejection \citep{poveda67}
mechanisms that explain runaway stars cannot produce ejection velocities larger than
200 - 300 km s$^{-1}$ for B-type stars \citep{leonard91, leonard93, portegies00,
davies02, gualandris04, dray05}.  The first HVS discovered, by comparison, has a
heliocentric radial velocity of +853 km s$^{-1}$ and a Galactic rest-frame velocity
of at least $+709\pm12$ km s$^{-1}$ \citep{brown05}, many times that needed to
escape the Milky Way.  Photometric follow-up shows that the object is a slowly
pulsating B main sequence star \citep{fuentes06}.  Only interaction with a MBH can
plausibly accelerate this 3 M$_{\sun}$ main sequence B star to such an extreme
velocity and distance ($\sim$110 kpc).

	HVSs are fascinating because they can be used to understand the nature and
environs of MBHs \citep[see the recent theoretical work of][]{gualandris05,
holley05, ginsburg06, levin06, perets06, baumgardt06, sesana06, bromley06,
merritt06, demarque07, ginsburg07, oleary07, gualandris07, kollmeier07}.  The
trajectories of HVSs also provide unique probes of the shape and orientation of the
Galaxy's dark matter halo \citep{gnedin05}.  Discoveries of additional HVSs
\citep{edelmann05, hirsch05, brown06, brown06b} are starting to provide suggestive
limits on the stellar mass function of HVSs, the origin of massive stars in the
Galactic Center, and the history of stellar interactions with the MBH.  Clearly, a
larger sample of HVSs will be a rich source for further progress on these issues.

	We have designed a successful targeted survey for new HVSs.  We use the 6.5m
MMT and the Whipple 1.5m Tillinghast telescopes to obtain radial velocities of faint
B-type stars, stars with lifetimes consistent with travel times from the Galactic
center but which should not otherwise exist in the distant halo.  Four earlier
HVS discoveries from this survey are published elsewhere \citep{brown06,brown06b}.  
Here, we present spectroscopic observations of 643 new HVS candidates.

	Our paper is organized as follows.  In \S 2 we discuss our target selection
and spectroscopic identifications.  In \S 3 we present evidence for stars ejected
from the Galactic center on bound orbits.  In \S 4 we show that the unbound HVSs
have a marginally anisotropic distribution on the sky.  We conclude in \S 5.  Our
new observations are listed in Appendix A.

\section{DATA}

\begin{figure}		
 \includegraphics[width=3.25in]{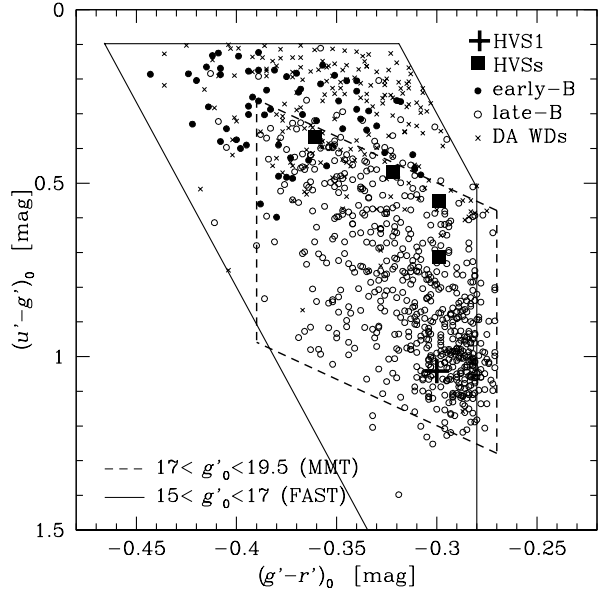}
 \figcaption{ \label{fig:ugr}
	Color-color diagram showing the target selection regions for the MMT 
({\it dashed line}) and FAST ({\it solid line}) samples, and the spectroscopic 
identifications for the combined set of 894 HVS candidates.  The brighter FAST 
sample ($15<g'_0<17$) targets bluer stars and finds a number of stars with 
B3 and B4 spectral types ({\it solid dots}).}
 \end{figure}

\subsection{Target Selection}

	\citet{brown06b} describe our target selection for candidate HVSs.  
Briefly, we use Sloan Digital Sky Survey (SDSS) photometry to select late B-type
stars by color. Figure \ref{fig:ugr} presents the HVS survey color-color selection.  
We use the 6.5m MMT telescope to observe every star with $17<g'_0<19.5$ in this 
color range.  We have observed 136 new HVS candidates and have completed the high
declination region of the SDSS Data Release 4 \citep[DR4,][]{adelman06}.  The MMT
survey is now 80\% complete across SDSS DR4 and covers an area of $\sim$5,000
deg$^2$ or 12\% of the entire sky.

	We have also carried out a new, complementary HVS survey targeting bright
stars with $15<g'_0<17$ and B-type colors in the SDSS.  We use the 1.5m Tillinghast
telescope and the FAST spectrograph \citep{fabricant98} to observe these bright HVS
candidates. We are able to survey efficiently much bluer stars with FAST because
contamination from white dwarfs is less problematic at bright magnitudes.  Figure
\ref{fig:ugr} shows the FAST color-color selection:  $0.1<(u'-g')_0<1.5$ and
$(10.67(g'-r')_0 + 3.5) < (u'-g')_0 < (10.67(g'-r')_0 + 5.07)$.  This region follows
the stellar sequence of B-type stars in the SDSS photometric system
\citep{fukugita96}.  Following \citet{brown06}, we impose a color cut
$-0.5<(r'-i')_0<0$ to reject objects with non-stellar colors.  We also exclude the
region of sky between $b<-l/5 + 50\arcdeg$ and $b>l/5-50\arcdeg$ to avoid excessive
contamination from Galactic bulge stars. There are 746 SDSS DR4 candidate B stars
with $15<g'_0<17$ in the selection region.  We have observed 514 stars, 69\% of this
total.  The average surface number density of bright HVS candidates is 0.13
deg$^{-2}$.  Thus we have surveyed an effective area of $\sim$4000 deg$^2$ for HVS
candidates with $15<g'_0<17$ with FAST.

\subsection{Spectroscopic Observations and Radial Velocities}

	New observations at the 6.5m MMT telescope were obtained with the Blue
Channel spectrograph on the nights of 2006 May 23-26 and 2006 June 19-20.  The
spectrograph was operated with the 832 line mm$^{-1}$ grating in second order and a
1.25$\arcsec$ slit.  These settings provided a wavelength coverage of 3650 \AA\ to
4500 \AA\ and a spectral resolution of 1.2 \AA.  Exposure times ranged from 5 to 30
minutes and were chosen to yield $S/N=15$ in the continuum at 4000 \AA.  Comparison
lamp exposures were obtained after every exposure.

	Observations at the 1.5m Tillinghast telescope were obtained with the FAST
spectrograph over the course of 19 nights between 2006 January 1 and 2006 July 21.  
The spectrograph was operated with a 600 line mm$^{-1}$ grating and a 2$\arcsec$
slit.  These settings provided a wavelength coverage of 3500 \AA\ to 5400 \AA\ and a
spectral resolution of 2.3 \AA.  Like the MMT observations, exposure times ranged
from 5 to 30 minutes and were chosen to yield $S/N=15$ in the continuum at 4000 \AA.  
Comparison lamp exposures were obtained after every exposure.

	Spectra were extracted using IRAF\footnote{IRAF is distributed
by the National Optical Astronomy Observatories, which are operated by the
Association of Universities for Research in Astronomy, Inc., under cooperative
agreement with the National Science Foundation.}
	in the standard way.  Radial velocities were measured using the
cross-correlation package RVSAO \citep{kurtz98}.  \citet{brown03} describe in detail
the cross-correlation templates we use.  The average uncertainty is $\pm11$ km
s$^{-1}$ for the B-type stars.

\subsection{Selection Efficiency and Completeness}

\begin{figure}		
 \includegraphics[width=3.25in]{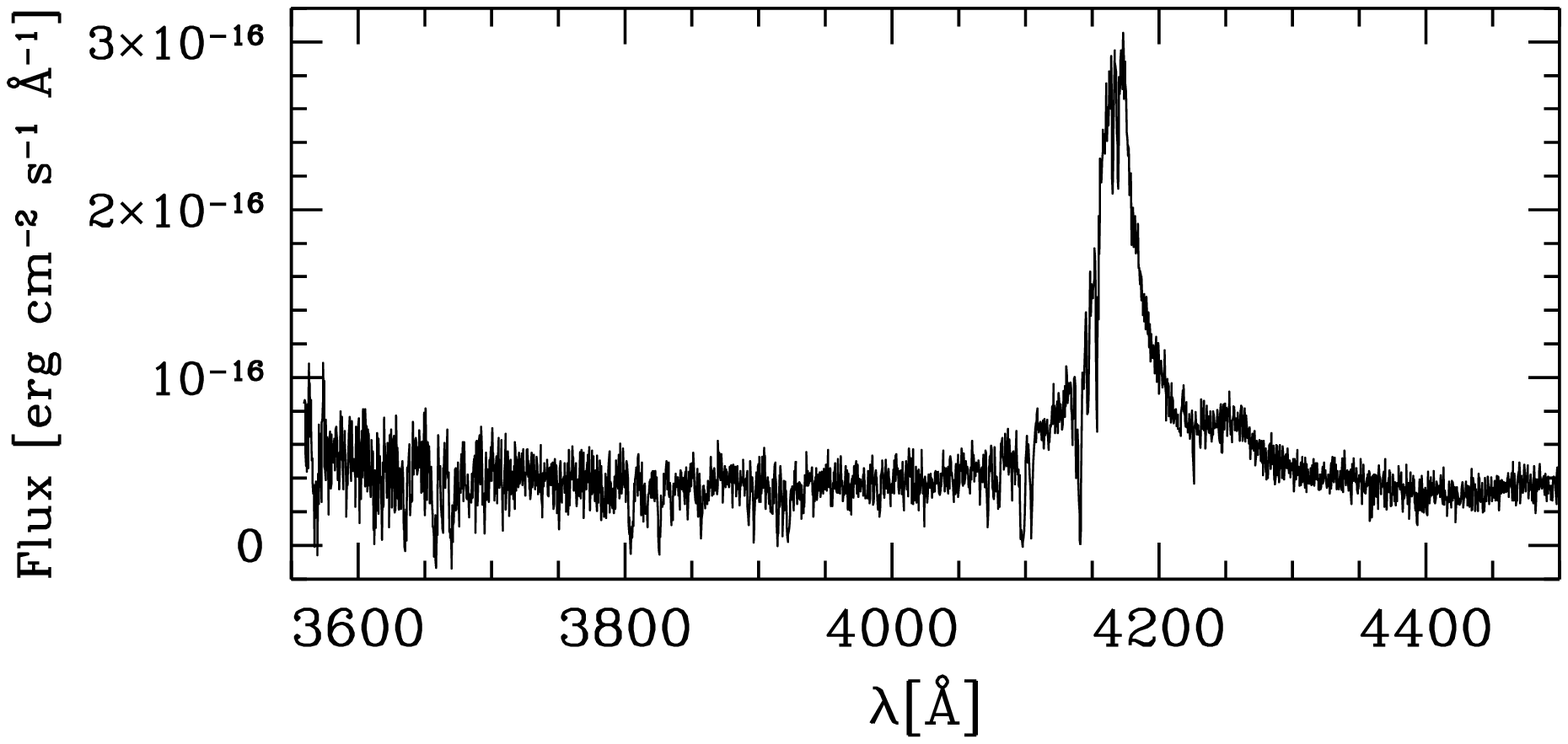}
 \figcaption{ \label{fig:qso}
	Spectrum of the $z=2.43$ quasar located at $15^{\rm h} 58^{\rm m} 51\fs85$,
$+22^{\rm d}21^{\rm m}59\fs9$ (J2000).}
 \end{figure}

	Our combined observations total 894 HVS candidates with $15<g'_0<19.5$, of
which 699 (78\%) are stars of B spectral type.  The remaining objects include 185 DA
white dwarfs, 3 other (DZ, DB) white dwarfs, 1 compact blue galaxy
\citep[see][]{kewley06}, 1 quasar, and 5 objects of uncertain identification,
possibly broad absorption line quasars.  The one obvious quasar is a previously
unidentified $g'=19.67$ object at $z=2.43$, located at $15^{\rm h} 58^{\rm m}
51\fs85, +22^{\rm d}21^{\rm m}59\fs9$ (J2000).  Figure \ref{fig:qso} presents its
spectrum.  The quasar has colors $(u'-g')_0=+0.80$ and $(g'-r')_0=-0.28$, unusually
red in $(u'-g')_0$ and blue in $(g'-r')_0$ compared to other quasars at similar
redshifts \citep{schneider05}.

	Figure \ref{fig:ugr} plots the spectroscopic identifications of the HVSs,
other B-type stars, and the white dwarfs.  Interestingly, we find stars with
spectral types as early as B3 in the bright FAST sample:  the solid dots in Figure
\ref{fig:ugr} are the 61 stars with spectral types B3 and B4.  If these early B-type
stars are main sequence stars, they are located at surprisingly large heliocentric
distances $30\lesssim d \lesssim60$ kpc.  We discuss the early B-type stars
further in \S 3.

	Our survey is now 100\% complete across the high declination-region of SDSS
DR4 located at $25<l<210\arcdeg$, $b>0\arcdeg$ in Galactic coordinates.  Curiously,
we find no new unbound HVSs in the completed region.  Our previous discovery rate
with the MMT was 1 HVS per $\sim$50 B-type stars \citep{brown06b}, yet our new set
of observations contains 245 (57 from the MMT) B-type stars in the complete region.
The absence of an unbound HVS in this large region suggests that the distribution of
HVSs on the sky may have structure.  We address the spatial distribution of HVSs in
\S 4.

	For the remainder of this paper, we consider a ``clean'' sample containing
560 stars selected only from the overlapping region of color-color space:
$-0.38<(g'-r')_0<-0.28$ and $2.67(g'-r')_0 + 1.30 < (u'-g')_0 < 2.67(g'-r')_0 +
2.0$.  Main sequence B stars in this selection region have masses of 3 - 5 M$_\sun$.  
We also consider the full sample of 699 B-type stars at all colors.  The full sample
contains earlier B-type objects from the FAST sample possibly including main
sequence stars with masses up to 7 M$_\sun$.  We emphasize that our photometric
selection is complete and can detect a B-type star at any velocity.

\section{BOUND HYPERVELOCITY STARS}

	The observed distribution of distances and velocities of our B-type stars
provides ample evidence that they are not main sequence run-aways.  Curiously, the
velocity distribution reveals an asymmetry of stars with large positive radial
velocities.  Because the large positive velocity outliers are very unlikely main
sequence run-aways, we argue that the stars represent a class of {\it bound} HVSs
ejected from the Galactic center.

\subsection{Background} 

	High Galactic latitude B-type stars were first reported by
\citet{humanson47}.  Early spectroscopic studies \citep{feige58, berger63,
greenstein66} showed that high latitude B-type stars are a mix of Population II
post-main sequence stars and Population I main sequence run-aways.  Accurate space
motions and distances are now known for nearby $d<10$ kpc B-type stars observed by
{\it Hipparcos}.  A detailed analysis by \citet{martin04} shows that 2/3 of the high
latitude B stars in the {\it Hipparcos} catalog are main sequence run-aways and 1/3
are evolved (mostly blue horizontal branch) halo stars.  \citet{martin04} finds that
all of the main sequence run-away B stars in the {\it Hipparcos} catalog have orbits
consistent with a disk origin.

	Our HVS survey probes much fainter and more distant B-type stars than
observed by {\it Hipparcos}.  Thus we cannot rely on proper motions or parallaxes.  
\citet{brown06b} discuss the nature of the late B-type stars and conclude that the
distribution of radial velocities and metallicities suggests that they are likely a
Galactic halo population of post-main sequence stars and/or blue stragglers.  We now
re-visit the possibility of main sequence run-aways in our sample.

	Theorists have shown that unbinding a stellar binary through supernova
disruption \citep{blaauw61} or binary-binary interactions \citep{poveda67} produces
a maximum ejection velocity of 200 - 300 km s$^{-1}$ \citep{leonard91, leonard93,
portegies00, davies02, gualandris04, dray05}.  The maximum velocity is set by the
escape velocity from the stellar surface.  For example, a contact binary containing
two 3 M$_{\sun}$ stars has a Keplerian orbital velocity of 240 km s$^{-1}$.  
Extracting a large ejection velocity from disrupting such a system is difficult.  
\citet{portegies00} considers supernovae unbinding binaries and finds that 90\% of
ejections have velocities between 0 and 100 km s$^{-1}$; only 1\% of 3 M$_{\sun}$
stars are ejected at 200 km s$^{-1}$.  \citet{davies02} perform similar calculations
for binaries that undergo a common envelope phase followed by a mass-transfer phase
that produces a type Ia supernova.  In such systems, Davies et al.\ find that 3
M$_{\sun}$ secondaries have typical ejection velocities $\sim$100 km s$^{-1}$ and
maximum ejection velocities of 250 km s$^{-1}$.  By comparison, the escape velocity 
of the Galaxy near the Sun is at least 500 km s$^{-1}$ \citep{carney88}. 

	Because main sequence run-aways travel on bound orbits, they spend most of
their time near the apex of their trajectories above the disk with small
line-of-sight velocities.  For example, a star ejected at 240 km s$^{-1}$ vertically
out of an infinite disk with surface mass density 95 M$_{\sun}$ pc$^{-2}$ reaches
an apex of $z=11$ kpc.  This simple calculation illustrates that main sequence
run-aways cannot simultaneously have large velocities $v>+275$ km s$^{-1}$ and large
distances $z>10$ kpc.  \citet{davies02} use a full Galaxy potential model and
compute orbits for 1000 run-away stars randomly ejected from their predicted
velocity distribution.  The vast majority of the \citet{davies02} run-aways are
located at $|z|<5$ kpc; at high Galactic latitudes $|b|>30\arcdeg$ the most distant
run-aways are found $\sim$25 kpc from the disk.

\subsection{The Asymmetric Velocity Distribution}

\begin{figure}		
 \includegraphics[width=3.25in]{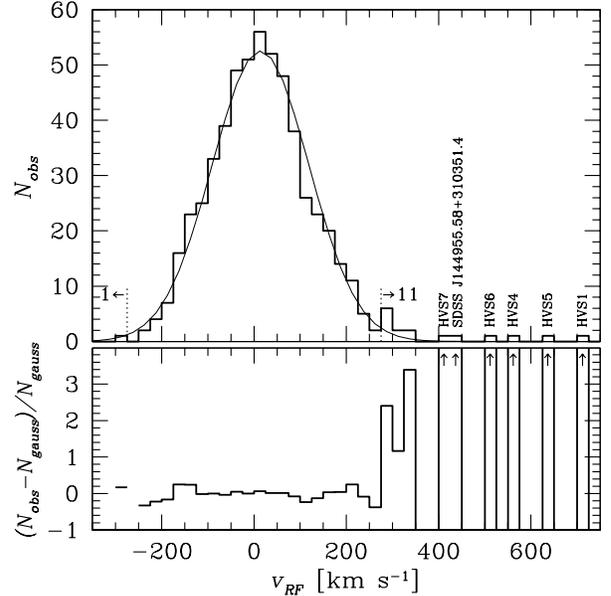}
 \figcaption{ \label{fig:hist}
	Galactic rest-frame velocity histogram of the clean sample of 560 B-type
stars ({\it upper panel}).  The best-fit Gaussian ({\it thin line}) has dispersion
$105\pm5$ km s$^{-1}$.  The lower panel plots the residuals of the observations from
the best-fit Gaussian, normalized by the value of the Gaussian.  In addition to the
unbound HVSs, there is a significant asymmetry of 11 positive velocity outliers
$+275<v_{rf}<450$ km s$^{-1}$ including SDSS J144955.58+310351.4.}
 \end{figure}

	Figure \ref{fig:hist} plots the distribution of line-of-sight velocities,
corrected to the Galactic rest-frame, for our clean sample of 560 late B-type stars
plus HVS1 \citep{brown05}.  It is apparent from Figure \ref{fig:hist} that the
velocities are well described by a Gaussian distribution.  We iteratively clip
$3\sigma$ outliers and calculate a $105\pm5$ km s$^{-1}$ dispersion and a $+14\pm4$
km s$^{-1}$ mean for the distribution.  We note that the full sample of 699 stars
has a statistically identical distribution.  The lower panel of Figure
\ref{fig:hist} plots the residuals of the observations from this Gaussian
distribution, normalized by the value of the Gaussian.  Stars with velocities
$|v_{rf}|<275$ km s$^{-1}$ show low-significance deviations from a Gaussian
distribution and probably constitute a halo population of post-main sequence stars
and/or blue stragglers \citep{brown06b}.  The highest-significant outliers are the
first HVS (HVS1), the four HVSs previously discovered in this survey (HVS4 - HVS7),
and a new object: SDSS J144955.58+310351.4.

	The possible HVS, SDSS J144955.58+310351.4, is a bright $g'_0=15.70\pm0.03$
star with an A1$\pm1.4$ spectral type and solar metallicity
[Fe/H]$_{W_k}=-0.3\pm0.6$.  A main sequence star of this type has heliocentric
distance $d\simeq17$ kpc \citep{schaller92}; an evolved blue horizontal branch (BHB)
star has $d\simeq7$ kpc \citep{clewley05}.  Located at $(l,b) = (48.7\arcdeg,
63.9\arcdeg)$, the star is high above the Galactic plane at $z\simeq15$ kpc and a
Galactocentric distance of $R\simeq17$ kpc (assuming $d=17$ kpc).  The star's
$+363\pm10$ km s$^{-1}$ heliocentric radial velocity corresponds to a minimum
velocity in the Galactic rest frame of $+447$ km s$^{-1}$ \citep[see][]{brown06b}.  
The star is listed with no proper motion in the USNOB1 \citep{monet03} catalog, and
thus is probably bound to the Milky Way.

	Curiously, we find a set of more distant stars with equally large, but
bound, velocities.  Figure \ref{fig:hist} shows a significant asymmetry in the wings
of the observed velocity distribution:  11 stars (excluding the 5 unbound HVSs) have
large positive velocities $+275<v_{rf}<+450$ km s$^{-1}$; only 1 star has a
comparable negative velocity at $v_{rf}=-286\pm12$ km s$^{-1}$.  This asymmetry was
first pointed out by \citet{brown06b} but is more significant in this larger sample.
There is a 0.00022 probability of drawing 11 stars with $+275<v_{rf}<450$ km
s$^{-1}$ and 1 star with $-450<v_{rf}<-275$ km s$^{-1}$ from a Gaussian
distributions with the observed parameters.  Integrating the wings of a Gaussian
with the observed parameters, we would expect to find 3.6 stars with
$+275<v_{rf}<450$ km s$^{-1}$ and 1.6 stars $-450<v_{rf}<-275$ km s$^{-1}$.  Thus
there is an excess of $\sim$7 stars with large positive velocity and no apparent
excess of stars with large negative velocity, significant at the $\sim$$3.5\sigma$
level.

	If the positive velocity outliers are main sequence stars, they are located
at large distances 20 - 80 kpc (see Table \ref{tab:bound}).  Yet the outliers cannot
be run-away stars, because the run-away mechanisms cannot produce simultaneously
large velocities and large distances.  If, on the other hand, the positive velocity
outliers are halo stars on radial orbits, we would expect to find equal numbers of
stars moving toward and away from us, contrary to observation.
	Compact binary systems may also produce outliers in the velocity
distribution.  But velocity outliers resulting from compact binaries should be
distributed symmetrically, again contrary to observation.  We note that neutron
stars are observed traveling at velocities of 1000 km s$^{-1}$, but such neutron
stars are the remnants of asymmetric supernova explosions, and are not B-type stars.

\subsection{Bound HVSs}

	One explanation for the significant excess of B-type stars traveling
$+275<v_{rf}<450$ km s$^{-1}$ is the HVS mechanism.  HVSs are ejected by three-body
interactions involving a MBH \citep{hills88}.  HVSs almost certainly come from the
Galactic Center.  The presence of a $3.6\times10^6$ M$_\sun$ MBH \citep{ghez05,
eisenhauer05} in the crowded Galactic center inevitably produces a ``fountain'' of
HVSs ejections.  The actual HVS ejection rate depends on the number of stars with
radial orbits in the MBH's so-called ``loss cone.'' If the MBH's loss cone is
replenished with stars scattered in by massive star clusters, molecular clouds, or
IMBHs \citep{perets06}, then the ejection rate can be orders of magnitude larger
than the original \citet{yu03} predictions.  Regardless of the exact rate, this
fountain picture results in a broad spectrum of HVS ejection velocities which
include bound and unbound orbits \citep{hills91, yu03, gualandris05, levin06, 
baumgardt06, perets06, sesana06, bromley06, oleary07}.  \citet{bromley06} recently
calculated ejection velocities for binaries containing 3-4 M$_{\sun}$ primaries
matched to our survey of B-type stars.


	Over the volume sampled by our survey, \citet{bromley06} predict comparable
numbers of HVSs ejected into bound and unbound orbits.  Our survey has discovered 4
HVSs on unbound orbits suggesting that $4\pm2$ of our excess $\sim$7 positive
velocity outliers are plausibly HVSs on bound orbits.  Although the predicted and
observed numbers of bound HVSs are statistically consistent, we note that additional
HVS mechanisms may be at work.  For example, \citet{oleary07} show that single stars
can be ejected by encounters with stellar mass black holes orbiting the central MBH.  
Such HVSs tend to have lower ejection velocities than HVSs ejected by the
\citet{hills88} mechanism, and thus may account for additional HVSs on bound orbits.

\begin{figure}		
 \includegraphics[width=3.25in]{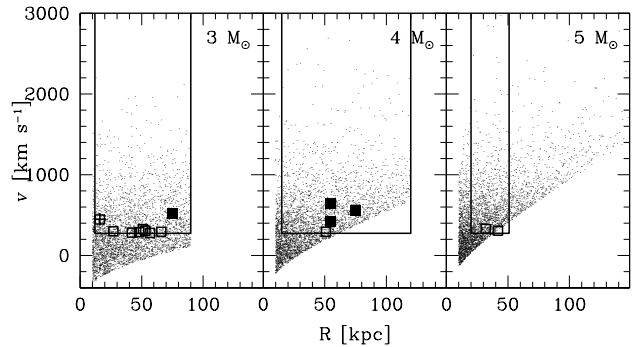}
 \figcaption{ \label{fig:hvs}
	Predicted velocity distributions of 3-5 $M_\sun$ HVSs with distance from the
Galactic center \citep{bromley06}.  We calculate distances for our B-type stars
assuming they are main sequence stars:  the four HVSs discovered by this survey
(solid squares) and the 11 positive velocity outliers $+275<v_{rf}<450$ km s$^{-1}$
(open squares) including SDSS J144955.58+310351.4 (open square with plus sign).}
 \end{figure}

	Figure \ref{fig:hvs} plots the \citet{bromley06} predicted velocity
distributions for main sequence HVSs ejected by the central MBH as a function of
Galactocentric radius.  HVS velocities increase with depth because only the most
rapidly moving HVSs survive long enough to reach large Galactocentric distances.  
The large squares in Figure \ref{fig:hvs} are our 4 HVSs (solid squares) and the 11
stars with $+275<v_{rf}<450$ km s$^{-1}$ (open squares).  We assume that the stars
are main-sequence stars, and use the observed colors and spectral types of the stars
to estimate stellar luminosity, distance, and stellar mass (see Table
\ref{tab:bound}).  The solid lines in Figure \ref{fig:hvs} indicate our survey
limits for main sequence stars of a given stellar mass.

	We evaluate the probability of drawing our observed sample from the
\citet{bromley06} predicted distributions.  We use the two-dimensional, two-sample
Kolmogorov-Smirnov test \citep{press92} to compare the observed and predicted
distributions of HVS velocity and distance.  First, we generate $10^5$ realizations
of the observations -- 4 HVSs plus 7 bound HVSs randomly drawn from the 11 high
velocity outliers -- and calculate the maximum difference in integrated probability
against the \citet{bromley06} distributions.  We use only the portion of the
\citet{bromley06} distributions that fall within our observational survey limits
(see Figure \ref{fig:hvs}).  Second, we randomly sample 11 objects from the
\citet{bromley06} distributions $10^5$ times and calculate the maximum difference in
integrated probabilities against the \citet{bromley06} distributions.  Again, we use
only the portion of the \citet{bromley06} distributions that fall within our
observational survey limits.  Finally, the likelihood is the fraction of time that
the probabilities of the synthetic data sets exceed the probabilities of the real
data sets.  The mean likelihood of drawing our observed sets of stars from the
predicted distributions is approximately 0.128, which neither strongly supports nor
strongly rejects the \citet{bromley06} model.  We conclude that the most plausible
explanation for the observed excess of positive velocity outliers is the HVS
mechanism ejecting stars from the Galactic center on bound orbits.

	We note that other stars in the literature may also be explained as bound
HVSs.  For example, the 5 M$_{\sun}$ main sequence B star HIP 60350 has a
heliocentric velocity of +230 km $s^{-1}$ and a full space velocity of 417 km
$s^{-1}$ based on {\it Hipparcos} measurements \citep{maitzen98}.  The star's large
radial motion $U=-352$ km $s^{-1}$ means that it originated from well inside the
solar circle, perhaps from the Galactic center.  

\subsection{Mystery of the Early B-type Stars}

	Our full sample of B-type stars contains many early-type stars at bright
magnitudes.  Four of the early B stars in the full sample are possibly bound HVSs
with $+275<v_{rf}<400$ km s$^{-1}$.  If they are main sequence stars, the four early
B stars have masses of 5 - 7 M$_{\sun}$ and they are located at distances ranging 30
- 60 kpc (bounded by the FAST survey limiting magnitudes).  We now ask whether the
number of early and late B-type bound HVSs are consistent with a standard initial
mass function.  Assuming the stars are main sequence stars, there are nine 3 - 5
M$_{\sun}$ stars in our sample located in the same volume as the four 5 - 7
M$_{\sun}$ stars: a ratio of 2.25.  A Salpeter initial mass function predicts a
ratio of 2.7, similar to what we observe.  However, the lifetimes of 5 - 7
M$_{\sun}$ main sequence stars are only 40 - 90 Myr \citep{schaller92}, a factor of
2 - 3 shorter than 3 - 5 M$_{\sun}$ stars.  The travel times of the four early B
stars, assuming $v_{rf}$ is their full space velocity, range from $\sim$100 to
$\sim$150 Myr from the Galactic center.  Thus if the early B stars are main sequence
stars ejected from the Galactic center, they do not have high enough velocities to
survive to their inferred distances.

	How can we reconcile the lifetimes of the early B-type stars with their
travel times from the Galactic center?
	It is possible that the early B-type stars are blue stragglers, main
sequence stars that have undergone mass transfer or mergers.  Blue stragglers are
usually associated with globular clusters, yet surveys of field BHB stars in the
halo find that half of the stars are in fact high surface-gravity blue stragglers or
A dwarfs \citep{norris91, preston94, wilhelm99b, brown03, clewley04, brown05b}.  
Because HVSs likely originate in tight binary systems, it is possible that some HVSs
experience mass-transfer from their former companion prior to being ejected from the
Galactic center.  A blue straggler origin is one explanation for HVS2, an 8
M$_{\sun}$ HVS with a main sequence lifetime otherwise inconsistent with travel time
from the Galactic center \citep{edelmann05}.

\begin{deluxetable}{lrcccccl}           
\tabletypesize{\scriptsize}
\tablewidth{0pt}
\tablecaption{POSSIBLE BOUND HVSs\label{tab:bound}}
\tablecolumns{7}
\tablehead{
  \colhead{Catalog ID} & \colhead{$l$} & \colhead{$b$} & \colhead{$v_{rf}$} &
  \colhead{type} & \colhead{$R$\tablenotemark{a}} & \colhead{$R$\tablenotemark{b}} \\
  \colhead{} & \colhead{{\small deg}} & \colhead{{\small deg}} &
  \colhead{{\small km s$^{-1}$}} & \colhead{} & \colhead{{\small kpc}} & \colhead{{\small kpc}}
}
	\startdata
SDSS J074950.24+243841.2 & 196.1 &  23.2 & +293 & B8 & 66 & 27 \\
SDSS J075055.24+472822.9 & 171.5 &  29.4 & +307 & B6 & 42 & 17 \\
SDSS J075712.93+512938.0 & 167.0 &  30.9 & +329 & B7 & 32 & 13 \\
SDSS J081828.07+570922.1 & 160.4 &  34.2 & +283 & B9 & 42 & 24 \\
SDSS J090710.08+365957.5 & 186.3 &  42.2 & +280 & B9 & 57 & 29 \\
SDSS J110224.37+025002.8 & 251.2 &  54.4 & +326 & A1 & 51 & 23 \\
SDSS J115245.91-021116.2 & 274.9 &  57.5 & +305 & A1 & 53 & 18 \\
SDSS J140432.38+352258.4 &  65.3 &  72.4 & +293 & B8 & 51 & 18 \\
SDSS J141723.34+101245.7 & 357.2 &  63.6 & +289 & A1 & 48 & 19 \\
SDSS J142001.94+124404.8 &   2.5 &  64.8 & +300 & B9 & 27 & 12 \\
SDSS J144955.58+310351.4 &  48.7 &  63.9 & +447 & A1 & 16 &  9 \\
        \enddata
\tablenotetext{a}{Galactocentric distance for a main sequence star of the observed 
spectral type.}
\tablenotetext{b}{Galactocentric distance for a BHB star of the observed color.}
 \end{deluxetable}

	It is also possible that the early B-type stars are main sequence stars
ejected elsewhere in the Galaxy by dynamical interactions with compact objects like
intermediate mass black holes (IMBHs).  However, as the mass of a black hole
decreases, the cross-section of HVS interaction decreases \citep[e.g.][]{hills88}.  
A stellar binary must come much closer to an IMBH before the gravitational tidal
energy exceeds the binding energy of the binary.  Thus an IMBH produces fewer HVSs
than a MBH in the same environment.  Unless IMBHs are ubiquitous in dense stellar
environments such as the centers of globular clusters, we expect that the overall
HVS rate is dominated by the MBH in the Galactic center.

\begin{figure*}		
 \centerline{\includegraphics[width=6.0in]{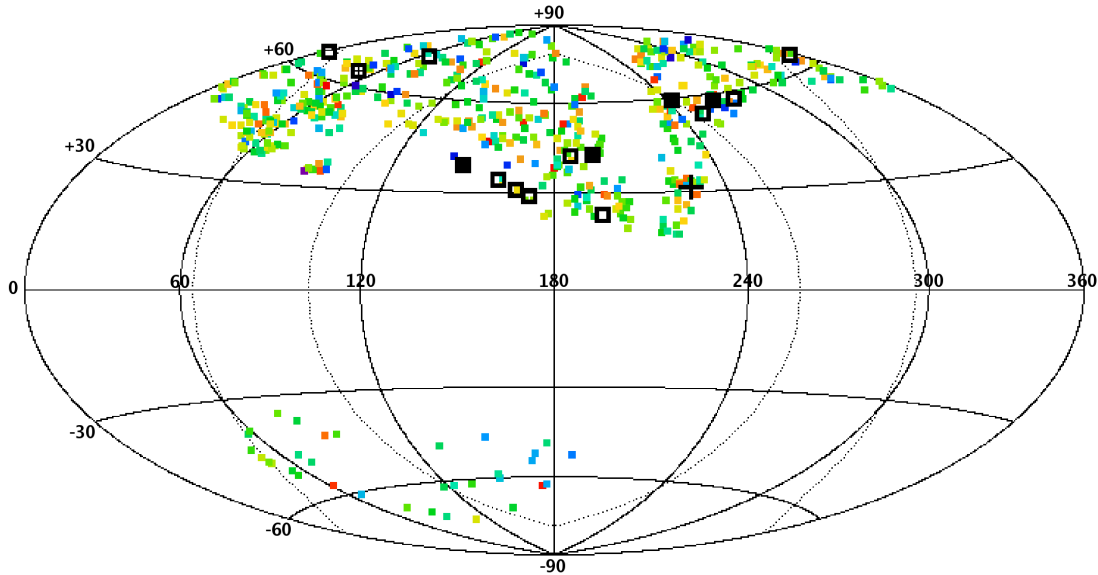}}
 \figcaption{ \label{fig:sky}
	Aitoff sky map, in Galactic coordinates, showing the clean sample of 560
B-type HVS candidates.  Radial velocities, in the Galactic rest frame, are indicated
by the color of the solid squares:  purple is -300, green is 0, and red is +300 km
s$^{-1}$.  The HVSs are completely off this color scale and are marked by:  HVS1
(plus), the four HVSs discovered by this survey (solid squares), and the 11 positive
velocity outliers $+275<v_{rf}<400$ km s$^{-1}$ (open squares) including SDSS
J144955.58+310351.4 (open square with plus sign).  Dotted lines are lines of
constant unbound HVS fraction calculated from the \citet{bromley06} models.}
 \end{figure*}

	A final possibility is that the early B-type stars are post main-sequence
stars.  \citet{demarque07} argue that the Galactic center contains many more old
dwarf stars than young B stars, and thus the MBH should eject many more BHB stars
than main sequence B stars.  \citet{davies05} and \citet{dray06} argue that the
Galactic center contains tidally-stripped evolved stars (in effect, BHB stars with
small hydrogen envelopes) that mimic B stars.  B-type BHB stars with hot effective
temperatures and small hydrogen envelopes spend most of their helium-burning
lifetimes at intrinsically faint luminosities \citep{yi97,dray06}.  Thus, if they
are BHB stars, the early B-type stars are located 10 - 20 kpc from the Galactic
center and have travel times ranging from 25 - 50 Myr.  The BHB progenitor lifetimes
easily exceed the inferred travel times, reconciling any difficulties with the MBH
ejection origin.  To test the post-main sequence explanation requires high
dispersion spectroscopy to determine the stellar nature of the bound HVSs.

\section{HVS SPATIAL ANISOTROPY}

	The distribution of HVSs on the sky is interesting because the spatial
distribution of HVSs is linked to their origin.  \citet{bromley06} show that
fraction of unbound HVSs increases with radial distance from the Galactic center
(see Figure \ref{fig:hvs}).  The fraction changes with depth because only the most
rapidly moving HVSs survive to reach large Galactocentric distances.  Because the
Sun is located 8 kpc from the Galactic center, our survey reaches approximately 16
kpc deeper towards the anti-center than towards the Galactic center.  Thus, if the
fountain model of HVSs ejected from the Galactic center is correct, we expect to
find more unbound HVSs towards the anti-center.

	We use the \citet{bromley06} models (Figure \ref{fig:hvs}) to calculate the
expected fraction of unbound HVSs in our survey volume as a function of position on
the sky.  The predicted fraction of unbound HVSs ranges from 53\% towards the
Galactic center to 63\% towards the anti-center.  The fraction does not vary much
because it is dominated by the survey volume at large distances.  The HVSs found in
this survey, however, are not located at distances beyond $\sim$80 kpc.  If we limit
the survey volume to the observed range of HVS distances, then the predicted
fraction of unbound HVSs ranges from 31\% to 46\%.  We now consider the
observations.

	Figure \ref{fig:sky} plots the distribution of our clean sample of 560
B-type stars across the sky in Galactic coordinates.  Color indicates the radial
velocity, in the Galactic rest frame, of the stars. The four unbound HVSs (solid
black squares) discovered by this survey are completely off the color scale; HVS1 is
marked by a plus sign.  The 11 possible bound HVSs are indicated by open squares;
SDSS J144955.58+310351.4 is plotted with an open square with a plus sign.  The
dotted lines in Figure \ref{fig:sky} are lines of constant HVS fraction calculated
from the \citet{bromley06} models.  We note that our survey is currently incomplete
in the SDSS equatorial slices, located at $l>210\arcdeg$ and $b<0\arcdeg$.


	It is striking that all of the unbound HVSs in our survey are located
towards the anti-center.  We randomize the observed velocities among the positions
of all 560 stars, and find a 0.064 probability of randomly drawing the 4 unbound
HVSs from our survey in the hemisphere $90<l<270\arcdeg$.  Thus the unbound HVSs
have a marginally anisotropic spatial distribution, being preferentially located in
the anti-center hemisphere with $\sim$2$\sigma$ confidence.  We note that the 3
unbound HVSs not from this survey, HVS1 \citep{brown05}, HVS2 \citep{edelmann05},
and HVS3 \citep{hirsch05}, are also located in the anti-center hemisphere
$90<l<270\arcdeg$.

	The 11 possible bound HVSs, on the other hand, appear more evenly
distributed across the sky:  6 bound HVSs are located $90<l<270\arcdeg$ and 5 bound
HVSs are located $|l|<90\arcdeg$.  To calculate the variation in unbound HVS
fraction, we consider only 7 bound HVSs (the excess number of high velocity outliers
in Figure \ref{fig:hist}) and split them 4/3 between the anti-center and
Galactic-center hemispheres, respectively.  The fraction of unbound HVSs is then
50$\pm$30\% in the anti-center hemisphere and 0$\pm$40\% in the Galactic-center
hemisphere.  Although small number statistics overwhelm our measurement, the unbound
HVS fraction appears consistent with the predictions of \citet{bromley06} and the
Galactic center ejection picture.

	There is another possible explanation for spatial anisotropy:  HVSs ejected
via single star encounters with a MBH binary are preferentially ejected in the
orbital plane of the MBH binary \citep{gualandris05, holley05, levin06, sesana06,
merritt06, baumgardt06}.  Recent N-body simulations for the case of an IMBH
inspiralling into a MBH, however, show that HVSs are ejected nearly isotropically
because the angular momentum vector of the IMBH rapidly changes on a $\sim$1 Myr
timescale \citep{baumgardt06}.  Moreover, our observed set of HVSs do not share
common travel times from the Galactic center.  Different IMBH inspiral events would
likely have different orbital histories, thus it seems unlikely that the observed
HVS spatial distribution can point back to a binary MBH origin.  A set of HVSs with
common travel times must be found in order to test the binary MBH origin.

\section{CONCLUSIONS}

	We discuss our targeted survey for HVSs, a spectroscopic survey of stars
with B-type colors.  We introduce a new component of the survey that samples
brighter and bluer stars using the FAST spectrograph.  Our combined set of
observations now contains 699 stars of B spectral type, 188 white dwarfs, and 1
previously unidentified $z=2.43$ quasar located at $15^{\rm h} 58^{\rm m} 51\fs85,
+22^{\rm d}21^{\rm m}59\fs9$ (J2000).

	We find no new unbound HVS, but we find evidence for a new class of {\it
bound} HVSs.  The velocity distribution of the B-type stars shows a significant
excess of $\sim$7 stars with large positive velocities $+275<v_{rf}<450$ km s$^{-1}$
and no apparent excess of stars with large negative velocity, significant at the
$\sim$$3.5\sigma$ level.  Neither halo stars on radial orbits, compact binary stars,
nor main sequence run-away stars ejected from the disk can explain simultaneously
the large velocities and distances of these stars.  The most plausible explanation
is that the positive velocity outliers are HVSs ejected from the Galactic center on
bound orbits.

	Our bright FAST survey contains a large number of early B-type stars, four
of which have velocities $+275<v_{rf}<450$ km s$^{-1}$.  These four possible bound
HVSs have main sequence lifetimes too short to survive to their inferred distances,
however.  One explanation is that the stars are post-main sequence stars.  
Establishing HVSs as main sequence or post-main sequence stars is important for
measuring the stellar mass function of HVSs and probing the types of stars that
orbit near the MBH.  Stellar rotation is a useful discriminant between rapidly
rotating main sequence B stars \citep{abt02,martin04} and slowly rotating BHB stars
\citep{peterson95,behr03}.  Metallicity is also a good discriminant between young
main sequence B stars and post-main sequence halo stars.  Echelle observations of
the brightest HVSs, underway now, will reveal the stars' true nature.

	A Galactic center origin predicts that unbound HVSs should be found
preferentially towards the anti-center as a result of survey selection.  We observe
that the unbound HVSs have a marginally anisotropic distribution at 2-$\sigma$
confidence.  Our estimates of unbound HVS fraction are consistent with the Galactic
center ejection picture.  Measuring the spatial distribution of a larger sample of 
HVSs may provide a strong test of the stars' origin.

	We note that full space motions for the HVSs will be known in a few years.
A {\it Hubble Space Telescope} program is measuring the stars' positions with 0.5
mas accuracies by registering background galaxies across Advanced Camera for Surveys
images.  The HVSs have predicted proper motions of 0.5 - 1 mas yr$^{-1}$, thus 
3$\sigma$ proper motion measurements are possible over a three year baseline 
\citep{gnedin05}.

	We are continuing our targeted HVS survey of faint B-type stars.  SDSS Data
Release 5 provides additional targets that we are now following-up with the MMT and
Whipple 1.5m telescopes.  Given our current discovery rate, we expect to find
another few unbound HVSs in the coming months.  Identifying HVSs throughout the
Galaxy will allow us to measure the mass function of stars in the Galactic center
and the history of stellar interactions with the central MBH.

\acknowledgements

	We thank M.\ Alegria, J.\ McAfee, and A.\ Milone for their assistance with
observations obtained at the MMT Observatory, a joint facility of the Smithsonian
Institution and the University of Arizona.  We thank H.\ Perets for helpful
correspondence.  This project makes use of data products from the Sloan Digital Sky
Survey, which is managed by the Astrophysical Research Consortium for the
Participating Institutions.  This research has made use of NASA's Astrophysics Data
System Bibliographic Services.  This work was supported by W.\ Brown's Clay
Fellowship and the Smithsonian Institution.

{\it Facilities:} MMT (Blue Channel Spectrograph), FLWO:1.5m (FAST Spectrograph)


\appendix
\section{DATA TABLE}

	Table 2 [on-line data table, attached as tab2.dat in the source] lists the
643 new observations from our HVS survey, excluding the extra-galactic objects.  
134 HVS candidates were observed with the MMT and 509 were observed with FAST (see
\S 2).  Table 2 includes columns of RA and Dec coordinates (J2000), $g'$ apparent
magnitude, $(u'-g')_0$ and $(g'-r')_0$ color, and our heliocentric velocity
$v_{helio}$.  The column WD indicates whether the object is a B-type star (WD=0) or
a white dwarf (WD=1).

	We note that $(u'-g')_0$ color correlates strongly with the observed
spectral type of our B-type stars.  Our spectral types are completely independent of
the SDSS photometry.  We classify spectral types based on \citet{oconnell73} and
\citet{worthey94} line indices as described in \citet{brown03}.  Typical uncertainty
is $\pm1.6$ spectral sub-types.  Figure \ref{fig:st} plots our spectral types and
the $(u'-g')_0$ color for our full sample of 699 B-type stars.  The solid line is
the best-fit relation:  spectral type $= 9.5(u'-g')_0 + 11.5$, where spectral type
10=B0, 15=B5, 20=A0, etc.  The spectral types of the stars in our HVS survey range
B3 - A2, with a dispersion around the best-fit relation of $\pm1$ spectral sub-types
or $\pm0.1$ mag in $(u'-g')_0$.

 \includegraphics[width=2.0in]{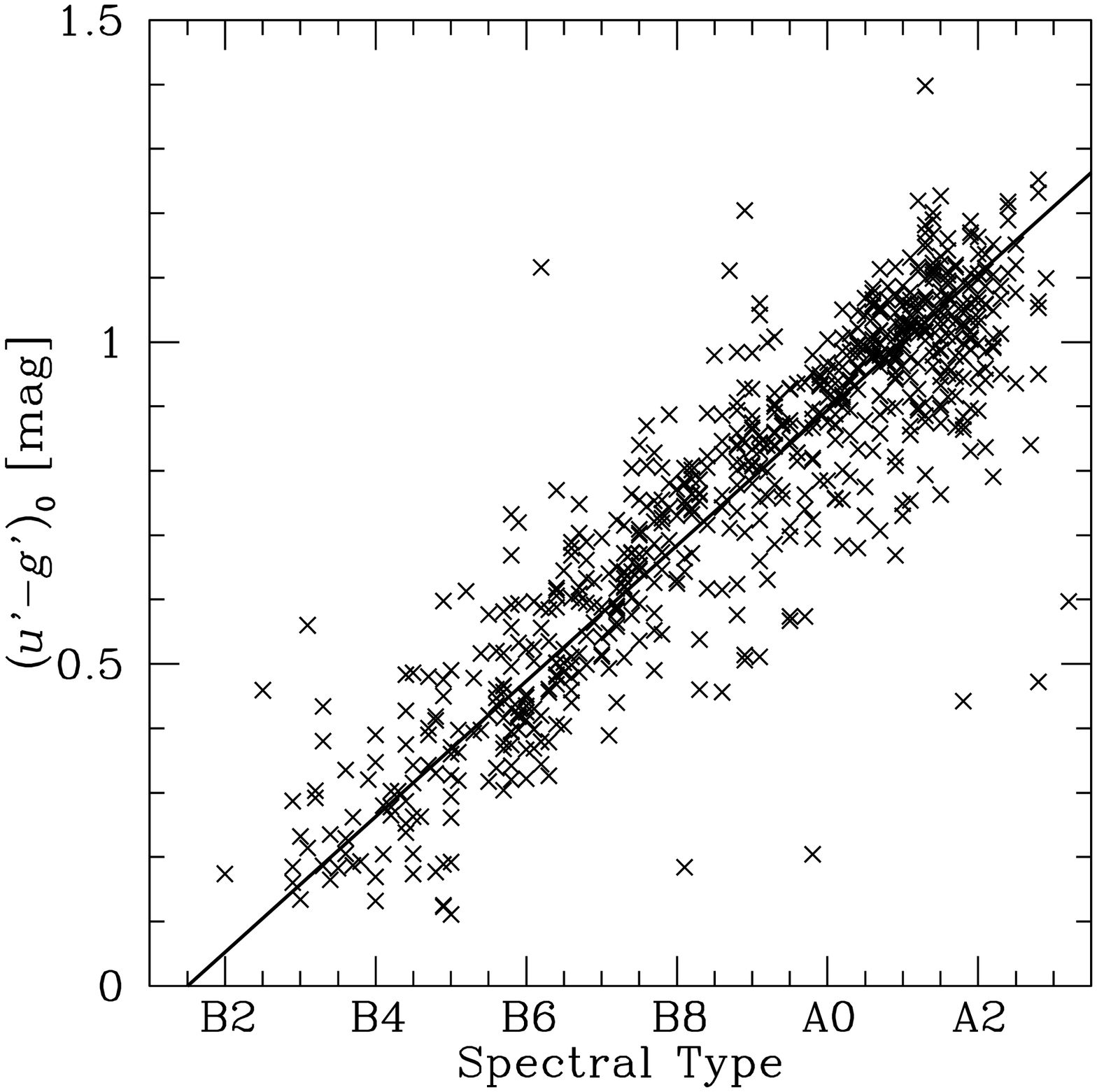}
 \figcaption{ \label{fig:st}
	Spectral types determined from our spectroscopy\\and SDSS $(u'-g')_0$ color
for the full sample of 699 B-type stars.}

\end{document}